\documentclass[10pt, conference,]{IEEEtran}
\IEEEoverridecommandlockouts
\usepackage{cite}
\usepackage{amsmath,amssymb,amsfonts}
\usepackage{booktabs}
\usepackage{algorithmicx}
\usepackage[noend]{algpseudocode}
\usepackage{graphicx}
\usepackage{textcomp}
\usepackage{xcolor} 
\usepackage{subfig}
\usepackage{algorithm}
\usepackage{tabularx}
\usepackage{makecell}
\usepackage{easyReview}

\def\BibTeX{{\rm B\kern-.05em{\sc i\kern-.025em b}\kern-.08em
    T\kern-.1667em\lower.7ex\hbox{E}\kern-.125emX}}

\begin{document}


\title{SkyOctopus: Enabling Low-Latency Mobile Satellite Network through Multiple Anchors\\
\thanks{This work was supported by the National Key R\&D Program of China under Grant No. 2023YFE0116600. The corresponding author is Wenjun Zhu. }
}
 \author{\IEEEauthorblockN{ Shaojie Su, Jiasheng Wu, Zijie Ying, Zhiyuan Zhao, Xiangyu Jia, Wenjun Zhu, Yue Gao}  \IEEEauthorblockA{Institue of Space Internet, Fudan University, China \\ School of Computer Science, Fudan University, China }}


\maketitle
\setreviewsoff
\begin{abstract} 

The rapid deployment of low earth orbit (LEO) satellite constellations has drawn attention to the potential of non-terrestrial networks (NTN) in providing global communication services. Telecom operators are attempting to collaborate with satellite network providers to develop mobile satellite networks, which serve as an effective supplement to terrestrial networks. However, current mobile satellite network architectures still employ the single-anchor design of terrestrial mobile networks, leading to severely circuitous routing for users and significantly impacting their service experience. 
To reduce unnecessary latency caused by circuitous routing and provide users with low-latency global internet services, this paper presents SkyOctopus, an advanced multi-anchor mobile satellite network architecture. SkyOctopus innovatively deploys traffic classifiers on satellites to enable connections between users and multiple anchor points distributed globally. It guarantees optimal anchor point selection for each user's target server by monitoring multiple end-to-end paths. We build a prototype of SkyOctopus using enhanced Open5GS and UERANSIM, which is driven by actual LEO satellite constellations such as Starlink, Kuiper, and OneWeb. We conduct extensive experiments, and the results demonstrate that, compared to standard 5G NTN and two other existing schemes, SkyOctopus can reduce end-to-end latency by up to 53\%.

\end{abstract}


\section{Introduction} \vspace{-0.5ex}

Nowadays, we are witnessing the rapid development of low earth orbit (LEO) satellite constellations, such as SpaceX's Starlink~\cite{starlink}, Amazon's Kuiper~\cite{Kuiper}, and OneWeb~\cite{OneWeb}. These satellite constellations, with their dense satellite distribution and inter-satellite links (ISLs), provide global internet and communication services while complementing terrestrial mobile networks in underserved areas in a cost-effective manner.

Meanwhile, as the developer of 5G, 3GPP has explicitly stated that non-terrestrial networks (NTN) will be an essential component of future 5G-Advanced and 6G networks~\cite{38811,38821,6Gwhite}. In practice, it has become a trend for telecom operators and satellite network providers to collaborate to build mobile satellite networks, as exemplified by partnerships such as T-Mobile with SpaceX~\cite{t-mobile} and AT\&T with AST~\cite{att}.

To ensure compatibility with existing terrestrial mobile networks, mobile satellite networks largely adopt the design of terrestrial mobile networks, moving only the access network to the satellite~\cite{38821}. Since the core network remains deployed on the ground and unchanged, each protocol data unit (PDU) session corresponds to a specific anchor point.
Given the global random access patterns of users, this single-anchor design poses significant challenges for the user data plane in satellite mobile networks. In such cases, user traffic transmission needs to traverse a fixed ground anchor point, resulting in circuitous routing and increased latency.

A natural solution to this issue is to deploy the anchor point onto the satellite nearest to the base station. By deploying anchor points on satellites and minimizing the distance between mobile network infrastructures, this approach aims to mitigate the long end-to-end latency caused by circuitous routing. However, considering the changes in network topology caused by the high-speed movement of satellites, users would face severe anchor point reselection issue, significantly impacting service continuity and incurring substantial reselection costs. Therefore, this solution is difficult to widely apply in real-world scenarios.

In this paper, we present SkyOctopus, an advanced multi-anchor mobile satellite network architecture. SkyOctopus supports the simultaneous existence of multiple anchor points within a single PDU session by using traffic classifiers deployed on satellites. It also employs a fine-grained selection strategy, which uses location-based criteria for the initial selection of anchor points and updates anchor point choices based on network conditions through continuous monitoring. Additionally, based on the correspondence between base stations and traffic classifiers, and parallelized signaling transmission, we design a new PDU session establishment process for SkyOctopus. Users can quickly establish PDU sessions without concern for the number of anchor points.

We construct a prototype of SkyOctopus using enhanced Open5GS~\cite{open5gs} and UERANSIM~\cite{ueransim}, driven by real LEO satellite ephemerides, including Starlink, Kuiper and OneWeb. Based on this prototype, we conducted extensive experiments, and the results indicate that SkyOctopus significantly reduces end-to-end latency by up to 53\% compared to the other three schemes and reduces session establishment time by 86\%.

Contributions of this paper can be summarized as follows:

\begin{itemize}

\item We for the first time expose the issue of high end-to-end latency caused by circuitous routing in emerging mobile satellite networks, which is essentially due to the design of single-anchor PDU session.

\item We propose SkyOctopus, an advanced mobile satellite network architecture that achieves low-latency global internet services through multiple anchor points and fine-grained anchor point selection strategy.

\item We construct a prototype of SkyOctopus and conduct comprehensive experiments to demonstrate its effectiveness in reducing end-to-end latency and its efficiency in terms of session establishment time.

\end{itemize}
The rest of this paper is structured as follows. Section II introduces the background of the problem and our motivation. Section III presents an overview of the proposed SkyOctopus architecture. Section IV provides detailed explanations of the three aspects of SkyOctopus. We conduct extensive experiments and analyze the results in section V. Section VI presents a review of related work in the field. Section VII discusses additional considerations and issues related to our work. Finally, Section VIII briefly concludes this work.

\section{Background} \vspace{-0.5ex} \label{chapect 2}

\subsection{PDU Session in Mobile Networks} \label{back:pdu}
\begin{figure}[]
	\centering
	\includegraphics[width=0.8\linewidth]{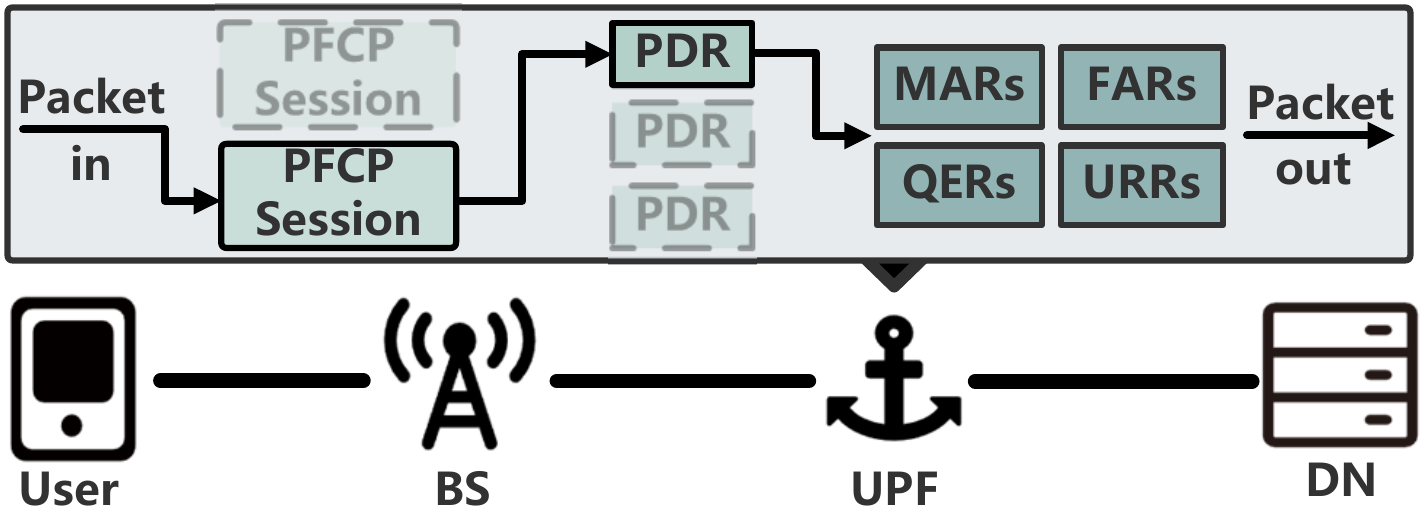}
	\caption{Process of the UPF handling input user traffic.}
	\label{Upf_uplane}
 \vspace{-0.5cm}
\end{figure}
In the 5G network, a PDU session refers to the logical channel between a user and a data network (e.g. Internet) through a specific base station and user plane function (UPF). The UPF manages the user session context and fixes the transmission path of user traffic, thus it is referred to as the anchor point of the PDU session in mobile networks. As the sole network function in the core network that handles user traffic, the UPF is responsible for executing all user plane policies according to packet detection rules (PDRs).

Specifically, Fig.~\ref{Upf_uplane} illustrates the process by which the UPF handles incoming user traffic~\cite{29244}. Taking the uplink direction as an example, upon receiving the user packet from the base station (BS), the UPF identifies the specified packet forwarding control protocol (PFCP) session to which the packet corresponds. Then, it selects the highest precedence PDR within the matching PDR of the PFCP session. Next, the UPF processes the data packet based on the associated rules specified by the selected PDR, including forwarding action rules (FARs), buffering action rules (BARs), QoS enforcement rules (QERs), and usage reporting rules (URRs). Finally, the packet is forwarded, with its direction determined by the matching FAR of the selected PDR.

3GPP has defined an intermediate UPF (I-UPF) that does not serve as an anchor point but is deployed as a traffic classifier between the base station and multiple UPFs~\cite{23501}. 
It achieves traffic classification by using different PDRs to match packets with various target IP addresses or data network names. These PDRs are often associated with different FARs, which forward the packets to different UPFs.
In terrestrial networks, the I-UPF is primarily used in private networks requiring high reliability (such as vehicular networks and smart factories) or multi-access edge computing (MEC) services to classify private and public network traffic.

PDU session with an I-UPF is established using an insertion-based process~\cite{23502}. For example, when a user moves into the service range of a specific MEC service, the core network inserts an I-UPF into the user's current PDU session and instructs the I-UPF to establish a connection with the additional UPF associated with the MEC service. When the user leaves the MEC service range, the I-UPF and the additional UPF are removed. 

\subsection{Mobile Satellite Networks} \label{mobile satellite network}

3GPP has defined two architectures for mobile satellite networks: the transparent mode (also known as bent-pipe) and the regenerative mode. In the transparent mode, the satellite acts as a transparent relay node between the user and the ground base station, whereas in the regenerative mode, base stations are deployed on the satellite, and user traffic can be forwarded through ISLs.

However, regardless of whether base stations are deployed on satellites, user traffic must be sent to the specific ground-based anchor point before being forwarded to the target server. Considering that users' access targets are randomly distributed globally, the anchor point often deviates from the path to the user's access server, causing significant detours. 
\begin{figure}[]
	\centering
	\includegraphics[width=0.8\linewidth]{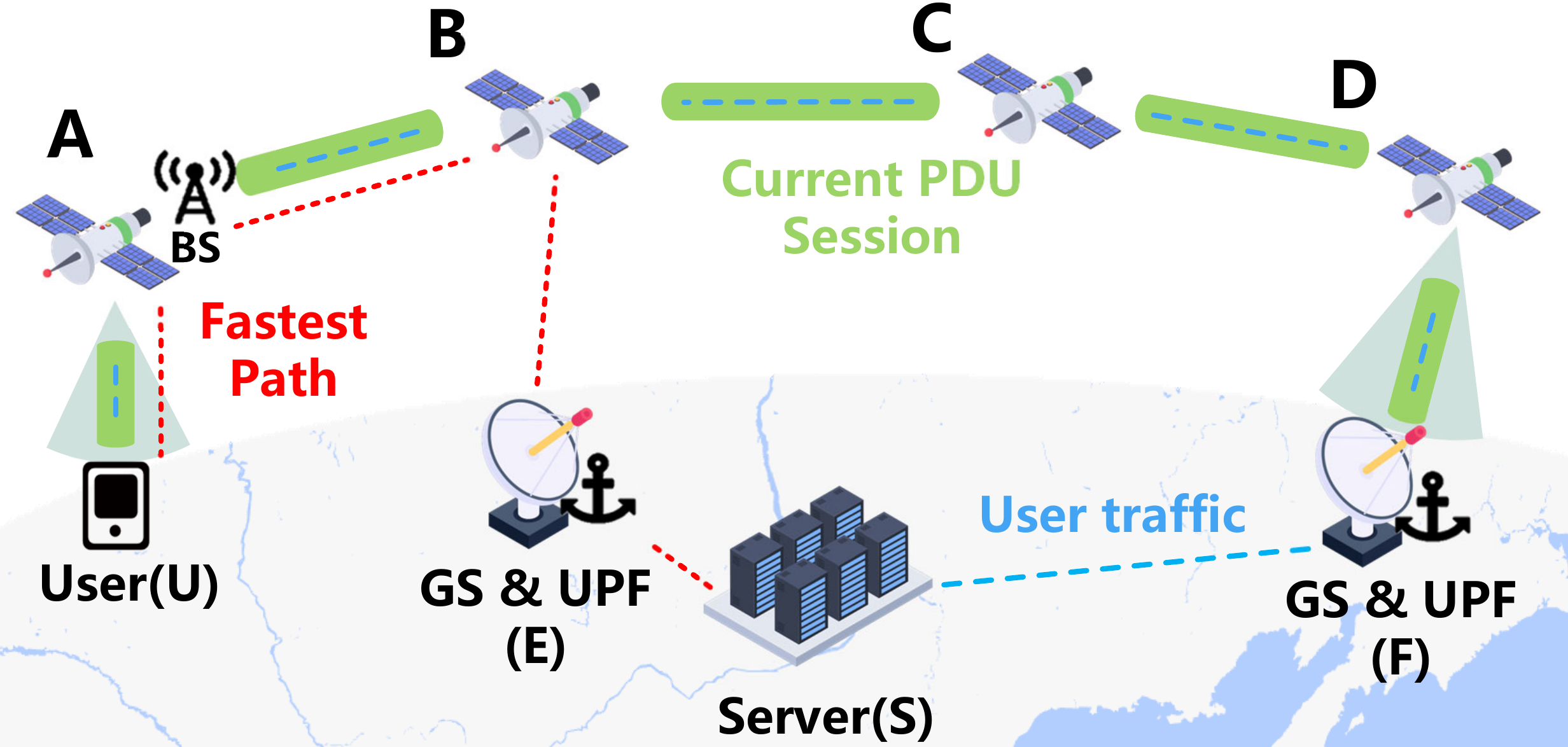}
	\caption{Circuitous routing in a mobile satellite network.}
	\label{route}
 \vspace{-0.5cm}
\end{figure}

Fig.~\ref{route} plots a typical example of circuitous routing in a mobile satellite network. For the current PDU session of UE $U$, the UPF at point $F$ serves as its anchor point. User traffic must first pass through the satellite network to reach the anchor point, and then proceed through the ground network to reach the server. Consider a scenario where the user accesses server $S$, then the user traffic follows the path $U - A - B - C - D - F - S$. It is evident that this path is not the fastest path in the network. In fact, $U - A - B - E - S$ is a more optimal path, which could significantly reduce latency.

As a specific example, consider the scenario of a user located in the Atlantic Ocean (42.2$^\circ$ N, 60.0$^\circ$ W) accessing a server in Paris through different paths within the Starlink constellation, as shown in Table I. When establishing a PDU session, the user selects the nearest ground station (GS) located in Ashburn, USA. Consequently, the user’s traffic is first transmitted via satellite to Ashburn and then through the terrestrial network to the server in Paris, with a total latency of 50.3ms. If the user selects the ground station in London as the anchor point, the total latency can be reduced to 26.8ms, a reduction of 44\%.

To address the issue of circuitous routing in mobile satellite networks, a straightforward method is to place the anchor point at the user satellite access point. 
This involves deploying a fully functional UPF on each satellite in regenerative mode. Users are provided with user plane services by the anchor point on the connected satellite, thereby avoiding detours caused by anchor points deviating from the shortest path. 

However, this design faces frequent anchor point reselection. When a base station handover occurs due to user or satellite movement, the anchor point is also reselected. Since the anchor point remains unchanged throughout the PDU session, this reselection means that users need to release the current PDU session and establish a new one. During the session reestablishment period, users experience an average service interruption of several hundred milliseconds, in addition to the interruption caused by the base station handover. More critically, session reestablishment can lead to the reassignment of the user's IP address \cite{23501}, causing interruptions in services that rely on connections. 
Considering the high-speed movement of LEO satellites, the reselection of anchor point occurs every 2-5 minutes~\cite{acchandover}, significantly impacting the service continuity for users. Therefore, this method is not a reasonable solution to the circuitous routing problem of mobile satellite networks.

\begin{table}[t!]
  \centering
   \caption{Latency comparison for different paths.}
   \vspace{-0.2cm}
    \begin{tabular}{c|c|c|c|c}
    \toprule
    \multicolumn{1}{c|}{Ground } & \multicolumn{1}{p{4.75em}|}{User to GS} & \multicolumn{1}{p{5.375em}|}{GS to Server } & \multicolumn{1}{p{4.915em}|}{Total Time} & \multicolumn{1}{p{5em}}{Time Saved} \\
    \multicolumn{1}{c|}{Station} & \multicolumn{1}{c|}{(ms)} & \multicolumn{1}{c|}{(ms)} & \multicolumn{1}{c|}{(ms)} & \multicolumn{1}{c}{(\%)} \\
    \midrule
    Ashburn & 4.8   & 42.5  & 50.3  & - \\
    \midrule
    London & 22.8  & 4     & 26.8  & 44 \\
    \bottomrule
    \end{tabular}%
  \label{tab:addlabel}%
  \vspace{-0.5cm}
\end{table}%

\section{Design Overview} \vspace{-0.5ex}

\textcolor{black}{Based on the above discussion, the essence of the circuitous routing problem in mobile satellite networks lies in the reliance on a fixed anchor point per PDU session and the difficulty of deploying the anchor point on satellites.}
This reliance directly leads to the issue where the anchor point is often not on the fastest path from the base station to the user’s target server, resulting in additional latency.
To address this, a natural approach is to expand the number of available anchor points in a single session and select different anchor points based on the actual target of the user's traffic to avoid detours.

In this paper, we propose SkyOctopus, a multi-anchor mobile satellite network architecture that enables users to have multiple available anchor points distributed globally within a single PDU session. In SkyOctopus, multiple UPFs are deployed as anchor points at ground stations. On one hand, by introducing the Satellite UPF (S-UPF), user traffic can be forwarded to different anchor points based on its target IP addresses. This design keeps the anchor points on the ground while moving the traffic classifiers to the satellite, thereby avoiding the circuitous routing problem and the frequent anchor point reselection issue caused by satellite mobility. 
\textcolor{black}{On the other hand, by redesigning the session establishment process, the S-UPF can establish connections with multiple anchor points simultaneously, avoiding the issue of prolonged session establishment times caused by the insertion-based process.}

However, there are two main challenges to applying this architecture. The first challenge is the anchor point selection problem. Although SkyOctopus allows for proactive routing selection, it is difficult to ensure reasonable anchor point selection to minimize end-to-end latency, considering the diversity of user targets and the mobility of satellites. The second challenge is the anchor point distribution problem, which involves determining the optimal locations for anchor points. Given the deployment and connection costs of UPFs, operators can hardly deploy anchor points without limitations. Therefore, it is necessary to strategically select their deployment locations given a fixed number of anchor points.

To address the first challenge, we propose a fine-grained anchor point selection strategy. The S-UPF uses PDRs based on the mapping of IP addresses to geographical locations to determine the initially chosen anchor point for users. Additionally, the path update mechanism ensures that the S-UPF can always select the optimal anchor point for users, even when network conditions change. The mechanism evaluates both intra-network and inter-network conditions to ensure users experience end-to-end low-latency access.

To tackle the second challenge, we analyze the anchor point deployment problem and prove it is an NP-hard problem. Based on this, we propose a greedy algorithm for selecting deployment locations for a fixed number of anchor points.

\section{Architecture Design} \vspace{-0.5ex}

In this section, we provide a detailed explanation of our proposed SkyOctopus architecture. First, we provide a comprehensive overview of SkyOctopus, along with detailed discussions on the modifications to network function, signaling, and protocol.

Then, we propose a fine-grained anchor point selection strategy, which includes initial location-based anchor point selection and the path update mechanism driven by both computation and measurement.

Finally, we discuss the anchor distribution problem and propose an algorithm to minimize latency by optimally selecting deployment locations for a fixed number of anchor points.

\begin{figure}[]
	\centering
	\includegraphics[width=0.8\linewidth]{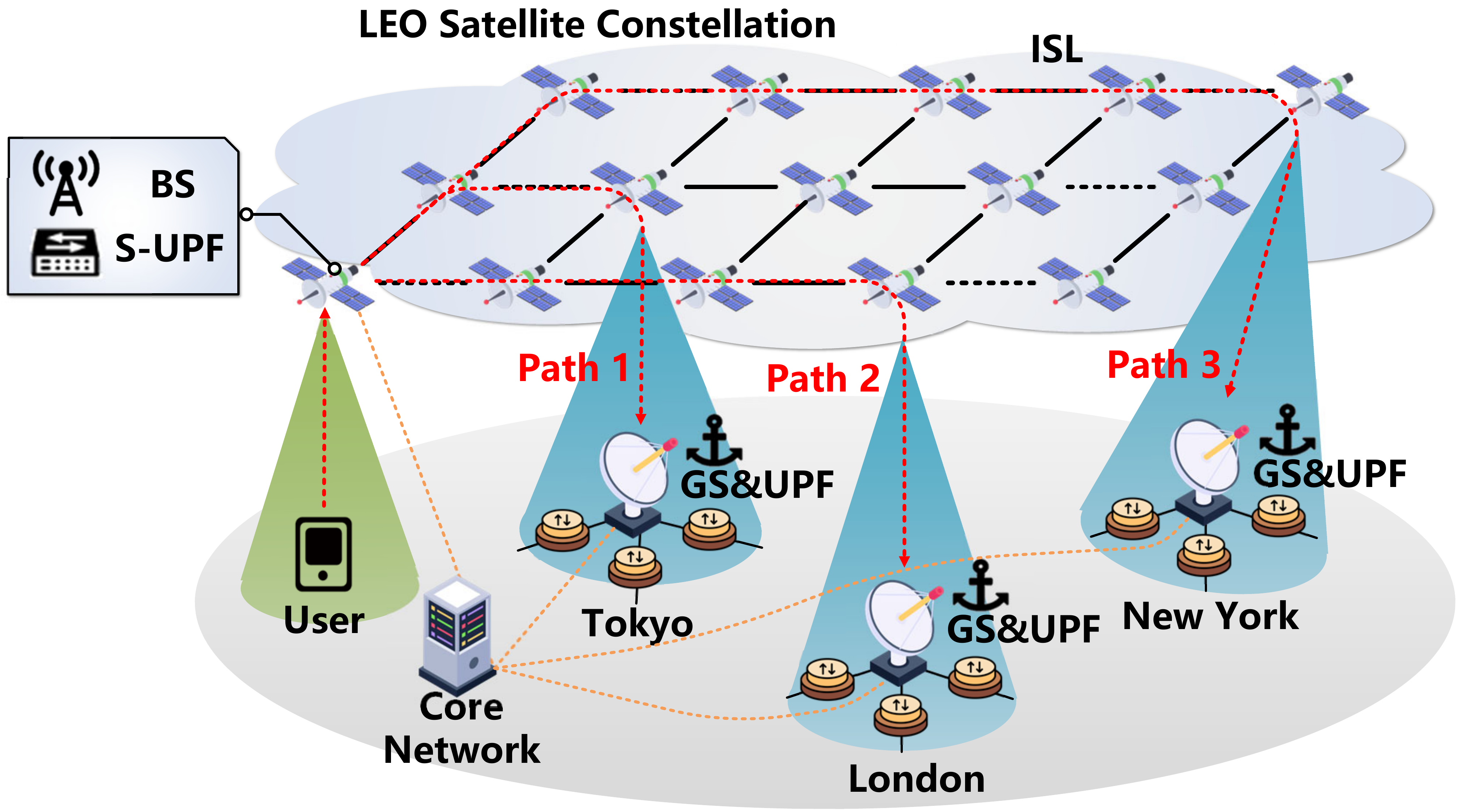}
	\caption{SkyOctopus architecture.}
	\label{SkyOctopus}
 \vspace{-0.5cm}
\end{figure}

\subsection{SkyOctopus Architecture}

To achieve low-latency global access for users, we propose SkyOctopus, an advanced multi-anchor mobile satellite network architecture, as shown in Fig.~\ref{SkyOctopus}. SkyOctopus operates within an LEO satellite constellation, where satellites establish ISLs with adjacent satellites both in the same orbit (front/rear) and in neighboring orbits (left/right). Given that these satellites share the same velocity and orbital altitude, the ISL topology remains stable.

SkyOctopus incorporates network components such as users, base stations, and the core network. Innovatively, SkyOctopus introduces a novel network function on the satellite, known as the Satellite UPF. This network function enables the existence of multiple anchor points (e.g., UPFs) on the ground within a single PDU session for a single user. The Satellite UPF is physically located on the satellite where the base station is situated, which helps further mitigate potential routing detours.

Below, we will detail the new network function, and outline the modifications made to the session establishment process in SkyOctopus.

\subsubsection{Satellite UPF}
In SkyOctopus, we introduce a new network function, named Satellite UPF (S-UPF). As a type of I-UPF, the S-UPF works as a traffic classifier operating on satellites. It splits user traffic from the base station on the satellite, forwarding it to different anchors based on the destination of the packets, and aggregates the traffic returning from multiple anchors back to the base station. 

On the other hand, since the S-UPF does not act as an anchor point, reselection of the S-UPF due to satellite movement does not result in the reestablishment of the user's current PDU session. Similar to the handover process of base stations, the new S-UPF is replaced into the current PDU session under the guidance of the core network, while the base station and anchor points still retain the session context. Therefore, users avoid the long service interruption time caused by anchor point reselection.

Compared to the standard NTN architecture, one significant innovation of SkyOctopus is that a single PDU session can include multiple anchor points. However, this feature presents a challenge for session establishment. Implementing the insertion-based session establishment process will cause the session establishment time to grow linearly with the number of anchor points.

In terrestrial mobile networks, a single session with I-UPF typically includes at most two or three anchor points due to the limited service range of private networks or MEC services. Thus, as described in \ref{back:pdu}, it employs an insertion-based process, where the core network sequentially adds each anchor point. However, in mobile satellite networks, where the number of potential anchor points far exceeds that in terrestrial scenarios, this method becomes impractical.

\begin{figure}[t]
	\centering
	\includegraphics[width=0.8\linewidth]{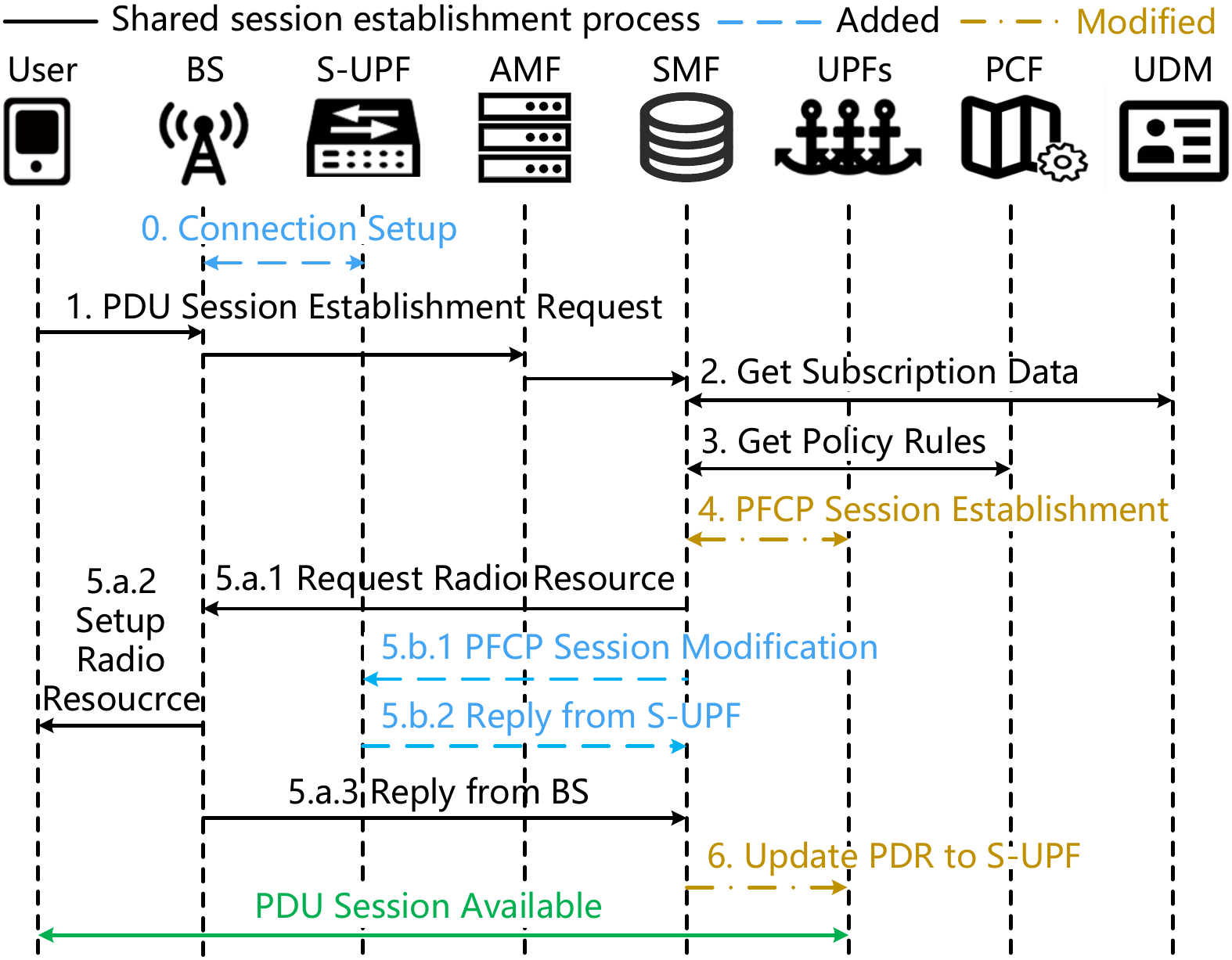}
	\caption{PDU session establishment process in SkyOctopus.}
 
	\label{session_establishment}
 \vspace{-0.5cm}
\end{figure}

We propose a new PDU session establishment process tailored for SkyOctopus, as illustrated in Fig.~\ref{session_establishment}. This process uses the correspondence between the base station and the S-UPF to directly establish connections with the anchor points, avoiding additional insertion operations. It also parallelizes connections across multiple anchor points, which together lead to a significant reduction in latency. The proposed PDU session establishment process can be summarized as follows.

\noindent\textbf{Step 0}. Before session establishment, the base station has already established a connection with the S-UPF on the same satellite based on the GTP-U protocol. In special cases such as link disruptions, the base station establishes a connection with the nearby S-UPF. Given that the ISL topology is fixed, the connection can remain stable. The S-UPF regularly reports the status of connected base stations to the core network in order to obtain the PDRs for these base stations.

\noindent\textbf{Step 1-3}. Similar to the traditional PDU session establishment process, the user initiates a PDU Session Establishment Request, which is transparently transmitted through the base station to the core network. Upon receiving the request, the SMF conducts necessary authentication for the user and retrieves the user's subscription data from the UDM  and policy rules from the PCF.

\noindent\textbf{Step 4}. Upon confirming the user's legitimate access, the SMF selects an anchor point responsible for allocating the IP address to the user based on the policy provided by the PCF. Differing from the traditional session establishment process, the SMF sends PFCP Session Establishment Request signaling to all anchor points rather than to a single anchor point. The signaling reply from each anchor point includes the tunnel endpoint identifier (TEID) for the GTP-U protocol that is prepared for establishing with the S-UPF. Additionally, the signaling reply from the anchor point responsible for assigning the IP address includes the user’s IP address. 

\noindent\textbf{Step 5}. The SMF then proceeds to send a radio resource request to the base station through the AMF (5.a.1) and transmits PFCP Session Modification to the S-UPF (5.b.1). 

Since the base station always establishes a connection with the default S-UPF and the S-UPF informs the core network of any changes in the connection, the SMF is aware of which S-UPF the base station is currently connected to. Therefore, the SMF can directly send the modification signaling to the corresponding S-UPF.


The base station assigns radio resources for the user (5.a.2), informs the user of the IP address allocated by the core network, and replies to the SMF with resource allocation success signaling (5.a.3). Subsequently, the S-UPF configures the TEIDs and PDRs accordingly based on the signaling and replies to the SMF with Modification Success signaling (5.b.2), providing the S-UPF's IP address and TEID list.

\noindent\textbf{Step 6}. The SMF logs the PDU session-related parameters replied to by the base station and verifies whether the TEID provided by the S-UPF matches the TEID replied by the UPF in Step 4. Furthermore, the SMF updates the anchor points' PDRs based on the S-UPF's IP to ensure that user traffic is correctly forwarded to the S-UPF instead of the base station.

The signaling in step 5.a and 5.b are transmitted in parallel. Given that the base station and the S-UPF are close, their transmission times to the core network are almost identical. Therefore, Step 5.b does not incur additional latency.

Through the above process, the user can simultaneously establish connections with multiple anchor points via the S-UPF, and the time required to establish a PDU session is decoupled from the number of anchor points. Additionally, the S-UPF, as a unique type of UPF managed by the core network, does not incur extra costs on the control plane and seamlessly integrates with mobile networks.

\subsection{Fine-Grained Anchor Point Selection}

\textcolor{black}{The availability of multiple anchor points expands the forwarding options for user traffic. Therefore, the key to reducing latency lies in selecting different anchor points for different target IP addresses within the user traffic. 
To address this issue, we design a fine-grained anchor point selection strategy for SkyOctopus, which includes location-based initial anchor point selection and a path update mechanism based on network status detection and computation.}

\subsubsection{Location-Based Anchor Point Selection} \label{geo_anchor_select}

In the proposed multi-anchor mobile satellite network, the S-UPF will determine the packet processing strategy based on the matching PDR, which specifies the transmission direction according to the user and their target IP. Given the vast number of possible user-target pairs---up to $10^{19}$ pairs for IPv4~\cite{ipv4out} and even more for IPv6---manually configuring PDRs for each user-server pair is impractical. To address this, we use location-based initial anchor point selection.

As an initial setup, anchor point selection should not depend on network status. A simple yet effective criterion for choosing anchor points is to select those that are as close as possible to the user’s target server. Compared to terrestrial public networks, satellite networks often have fewer hops, which provides advantages in terms of experimental performance and stability for the same transmission distance. Existing statistics indicate that current commercial LEO satellite networks outperform public networks in latency~\cite{starlinkPerformance}, and this advantage is expected to increase with the expansion of ground station numbers and advancements in the satellite industry.

On the other hand, it is possible to infer the geographical location based on the target IP address. Although IP addresses themselves do not contain geographical information, their distribution typically follows a geographic hierarchy. By examining the hierarchical arrangement and distribution details of IP addresses, one can infer the geographical location of a device. \textcolor{black}{A common approach is to use the IP geolocation database maintained by data providers such as GeoIP~\cite{geoip}}.

Considering the above points, we design the process for location-based anchor point selection as follows. Based on the geographical location corresponding to global IP addresses, we assign them to different anchor locations according to their distance. For each group, a PDR list is created based on target IP addresses, with subnet merging used to reduce the number of generated PDRs. The list is deployed as initially built PDRs in the S-UPF.

\begin{figure}[]
	\centering
	\includegraphics[width=0.8\linewidth]{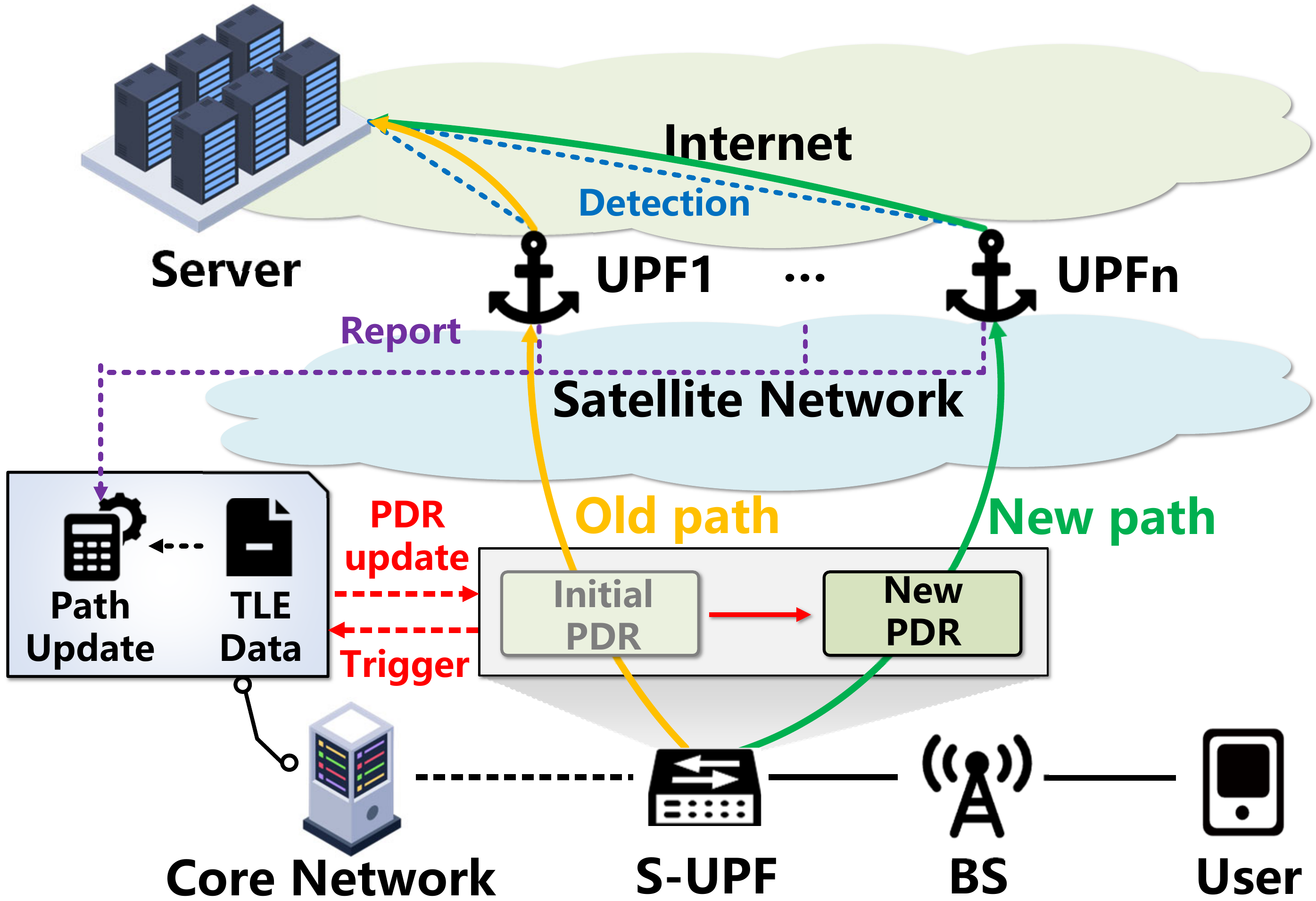}
	\caption{Work stage of the anchor point selection strategy.}
	\label{anchor_point_selection}
 \vspace{-0.5cm}
\end{figure}

\subsubsection{Path Update} 

Using location-based anchor point selection, we will choose an anchor point for user traffic that is most likely close to the target server. However, the initially chosen anchor point may not always be the optimal choice. First, location prediction based on IP addresses is not always accurate due to factors such as dynamic IP assignment and proxy server usage. Secondly, considering the complexity of network conditions and paths, the nearest anchor point to the server is not necessarily the best choice. To address these factors, we design a path update mechanism to ensure that users always achieve optimal end-to-end latency.

The mechanism evaluates both intra-network and inter-network conditions to ensure users experience end-to-end low-latency access. For intra-network evaluation, latency is primarily dominated by inter-satellite and satellite-to-ground propagation delays. Therefore, we use ephemeris data to calculate the latency between S-UPFs and anchor points. On the other hand, for inter-network evaluation, latency is difficult to predict due to the heterogeneity of Internet infrastructure and its random distribution. Consequently, we utilize ICMP~\cite{icmp} to observe the latency from each anchor point to the target server. Based on the calculations and observations of both internal and external network conditions, the core network computes the anchor point corresponding to the shortest latency path and notifies the S-UPF.

\subsubsection{Working Stages}
Fig.~\ref{anchor_point_selection} illustrates the working stages of the anchor point selection strategy. For a given user, when a packet with a specific destination IP first appears and matches the initial PDR in the S-UPF, it will be forwarded to the anchor point determined by the location-based anchor point selection.

Simultaneously, the path update mechanism is triggered. First, the S-UPF sends the IP address as a signaling parameter to the core network, which instructs anchor points to send probe messages to detect their latency to the target IP. Based on the output of the computational module and the latency returned by the anchor points, the core network calculates the latency from the S-UPF to the target server via each anchor point. The core network then creates a PDR for the S-UPF, indicating the optimal anchor for the target IP with a higher matching priority than the initially built PDR. The path update mechanism remains active, and if the latency through an alternative anchor point is superior to the current one, the core network initiates a PDR modification instead of creation.

\subsection{Anchor Point Distribution}

\textcolor{black}{In addition to anchor point selection, the distribution of anchor points is also a crucial factor affecting overall network latency. With a poor distribution, even the optimal path selection might result in high latency (e.g., if all anchor points are concentrated in Asia while a user tries to access services in the Americas). In this section, we will discuss how to strategically deploy anchor points globally to minimize the overall average network latency.}
\subsubsection{\textcolor{black}{Formal Description}}

 $ Sat = \{sat_1,sat_2,...sat_m\}, G= \{g_1,g_2,...g_n\}$ 
  \textcolor{black}{is denoted as }a satellite set contains $m$ satellites, and the ground station set contains $n$ ground stations. Let \textcolor{black}{$AP = \{ap_1,ap_2,...ap_h\} \in G$} denotes the deployment locations of $h$ ground anchors selected from the set of ground stations. $U = \{ u_1, u_2, \ldots, u_p \}$ and $S = \{ s_1, s_2, \ldots, s_q \}$ \textcolor{black}{denote respectively} all the $p$ users of the satellite network and the $q$ servers providing network services. $p(u,s)$ refers to the probability that user $u$ will use service $s$, where $u \in U$ and $s \in S$. We consider a propagation path from user $u$ to service $s$ through an anchor point $ap$. Using $ap$ as a boundary, the network can be divided into internal and external parts, with latencies represented by $L_{\text{in}}(u, ap, Sat)$ and $L_{\text{out}}(ap, s)$, respectively. 

When user $u$ needs to access server $s$, our proposed scheme can find an $ap$ that the total latency is minimized. The $h$-anchor distribution problem aims to determine the optimal distribution of $h$ anchor points that minimizes the average user-to-service latency across the network, which can be formulated as:
\begin{equation} \label{eq:1}
 \mathop{\min} \quad {\mathop{\sum}_{u \in U}\sum_{s \in S}{p(u,s)(\mathop{\min}_{ap\in{AP}}({L_{\text{in}}(u, ap, Sat)+ L_{\text{out}}(ap, s)})}}.
\end{equation} 
Let $x_{i,j,k}$ be a binary decision variable indicating the selection of anchor point for user and server, where $x_{i,j,k} = 1 $ means that $g_i$ is the chosen anchor point for $u_j$ and $s_k$. Let $y_i$ represent the deployment of anchors at ground stations, where $y_i = 1$ indicates that $g_i$ is selected for an anchor point's deployment. The definition of $x$ and $y$ can be can be expressed as: 
\begin{equation} \label{eq:2}
x_{i,j,k} = \left\{  
             \begin{array}{ll}  
             1, & \textit{$g_i$ is the chosen anchor point for $u_j$ and $s_k$}  \\  
             0, & otherwise\\    
             \end{array}  
\right.  
\end{equation} 
and 
\begin{equation} \label{eq:3}
y_{i} = \left\{  
             \begin{array}{ll}  
             1, & g_i \in AP  \\  
             0, & otherwise\\    
             \end{array}  .
\right.  
\end{equation} \label{eq:3_5}
To simplify the equations, let $PL_{ap}$ represent the expected latency of $u$ and $s$ via $ap=g$, defined as:
\begin{equation} \label{eq:4}
PL_{g,u,s} = p(u,s)({L_{\text{in}}(u, g, Sat)+ L_{\text{out}}(g, s)}).
\end{equation} 
Considering Eq. (\ref{eq:1}) - (\ref{eq:4}) comprehensively, we can transform the h-anchor distribution problem into a constrained optimization problem as:
\begin{align}
\min \quad  &\mathop{\sum}_{j=1}^{p}\mathop{\sum}_{k=1}^{q}\mathop{\sum}_{i=1}^{n} x_{i,j,k}{PL}_{g_i,u_j,s_k} \label{eq:5}\\
\text{s.t.} &\quad x_{i,j,k} \leq y_i, \ \forall i \in [n], j \in [p], k \in [q], \label{eq:6}\\
& \mathop{\sum}_{i=1}^{n} x_{i,j,k} = 1, j \in [p], k \in [q], \label{eq:7}\\
& \mathop{\sum}_{i=1}^{n} y_{i} = h. \label{eq:8}
\end{align}
The objective function (\ref{eq:5}) is derived from Eq. (\ref{eq:1}) based on the definition of $x$. Constraint (\ref{eq:6}) states that a ground station can be chosen as an anchor point only if an anchor point is deployed at that station. Constraint (\ref{eq:7}) ensures that for each user-service pair, there is exactly one chosen anchor point. Constraint (\ref{eq:5}) limits the size of the anchor point set $AP$.

\noindent\textbf{Theorem 1. The \textbf{\textit{h}}-anchor distribution problem is NP-hard.}

\noindent \textbf{Proof.} Let $t = pk + j$, where $k \in (1, q)$ and $j \in (1, n)$. Use $o_t \in O$ to denote the combination of $u_k$ and $s_j$. If we define the distance between $g_i$ and $o_t$ as $d_{{g_i},o_t} = PL_{i,j,k}$, this problem can be considered as a $k$-median problem but not in metric space. It can be reduced to the typical $k$-median problem. Since the $k$-median problem is NP-hard~\cite{kmedian}, the $h$-anchor problem is also NP-hard.

\subsubsection{Greedy Algorithm for $h$-anchor Distribution Problem}
To solve the above problem in polynomial time, we propose a greedy algorithm to obtain an approximate solution. In general, we iteratively exclude the ground station that has the least impact on the objective function. We continue the process until $h$ ground stations remain.

\begin{algorithm}[t]
\caption{Greedy algorithm for $h$-anchor distribution}  \label{alg:greedy} 
{{
    \begin{algorithmic}[1]
    \Require $G$, $h$
    \Ensure $G_c$
    \State Initialize $G_c = G$
    \While{$|G_c| > h$}
    \State $g_m = \arg\min{Cost(G_c-\{g\})}$
    \State $G_c= G_c-\{g\}$
    \EndWhile
    \State return $G_c$
    \end{algorithmic}
    }
    }
\end{algorithm}

The pseudo code of the proposed algorithm is shown in Algorithm \ref{alg:greedy}. Let $Cost(G_c)$ be the value of the objective function (\ref{eq:5}) for $AP=G_c$. We first initialize the selected ground node set $G_c = G$ (line 1). Next, we traverse $G_c$. For each point $g$, we compare the $Cost(G_c-\{g\})$ values and find the one with the minimum value, denoted as $g_{m}$ (line 3). We consider this ground node to have the least contribution to the objective function, and thus exclude it from $G_c$, resulting in $G_c = G_c - \{g_{m}\}$ (line 4). We continue this process until we obtain the set $G_c$ with $h$ elements (line 2), and we get $G_c$ as the selected $h$ ground station for anchor point deployment (line 5).


%

\section{Experiments}  \vspace{-0.5ex}

\subsection{Experimental Setup}

\noindent\textbf{Satellite Constellation}: We simulate the orbital trajectories of LEO satellites using public information from three different constellations. The Starlink constellation consists of 1,584 satellites in 72 orbitals~\cite{starlinkLEO}, while the Kuiper constellation consists of 1296 satellites in 36 orbitals~\cite{kuiperLEO} and the OneWeb constellation consists of 636 satellites in 12 orbitals~\cite{onewebLEO}. Based on these constellations, we have built a platform for simulating real-trace dynamics of LEO satellite constellations using skyfield~\cite{skyfield}.

\noindent\textbf{System Prototype}: Driven by the above platform, we have built a prototype using Open5GS~\cite{open5gs} and UERANSIM~\cite{ueransim} for SkyOctopus. Open5GS is a widely utilized simulator for core network implementation in 5G, while UERANSIM is employed for implementing users and base stations. Following the 5G standard protocols and signaling procedures, we have made modifications to Open5GS to realize the S-UPF and the proposed PDU session establishment process. The built prototype operates on a commercial laptop with a 2.7 GHz CPU core and 24 GB RAM.

\noindent\textbf{Ground Station and User Traffic}: 
We use 40 servers from AWS's 29 available zones~\cite{AWS} as ground stations. Among these, 20 ground stations will be selected as anchor points, which means $h = 20$. 
For user traffic, we randomly generate 100 users navigating through different locations in the Atlantic Ocean. Additionally, we select 25 target servers randomly from the top 50 most visited websites globally~\cite{website}. 

\noindent\textbf{Mobile Satellite Network Schemes}: We compare SkyOctopus with the following three mobile satellite network schemes to demonstrate its effectiveness in reducing latency.
\begin{itemize}
\item \textbf{Standard} refers to the standard 5G NTN architecture, which is described in \ref{mobile satellite network}. In this architecture, base stations are deployed on satellites and ISLs are available. The core network assigns a unique anchor point to the user based on their registered area and maintains this assignment for the duration of the PDU session.

\item \textbf{Standard-GS} refers to a scheme that selects the nearest ground station, meaning the core network assigns the closest anchor point to the user. User traffic is quickly transmitted to the corresponding ground station after passing through the base station on the satellite and then reaches the target server via the public network.

\item \textbf{Standard-SAT} refers to a scheme in which anchor points are deployed in different clusters on satellites, and the user is assigned an anchor point belonging to the cluster of the satellite access point, inspired by~\cite{skycastle}. This scheme reduces latency by deploying anchor points on satellites and shortening the distance between the base station and the anchor point.

\end{itemize}

\noindent\textbf{Anchor Point Distribution}: We compare the proposed greedy algorithm with the following two algorithms to demonstrate its effectiveness in selecting anchor point distribution.
\begin{itemize} 
\item \textbf{K-means} refers to the anchor distribution algorithm based on K-means. The algorithm clusters all available ground stations using K-means and further selects the center of each cluster as the deployment location for anchor points. 

\item \textbf{Random} refers to the algorithm that selects the anchor points randomly from all available ground stations. 
\end{itemize}

\begin{figure*}[t]
\centering
   \subfloat[Starlink]{\includegraphics[width=0.33\textwidth]{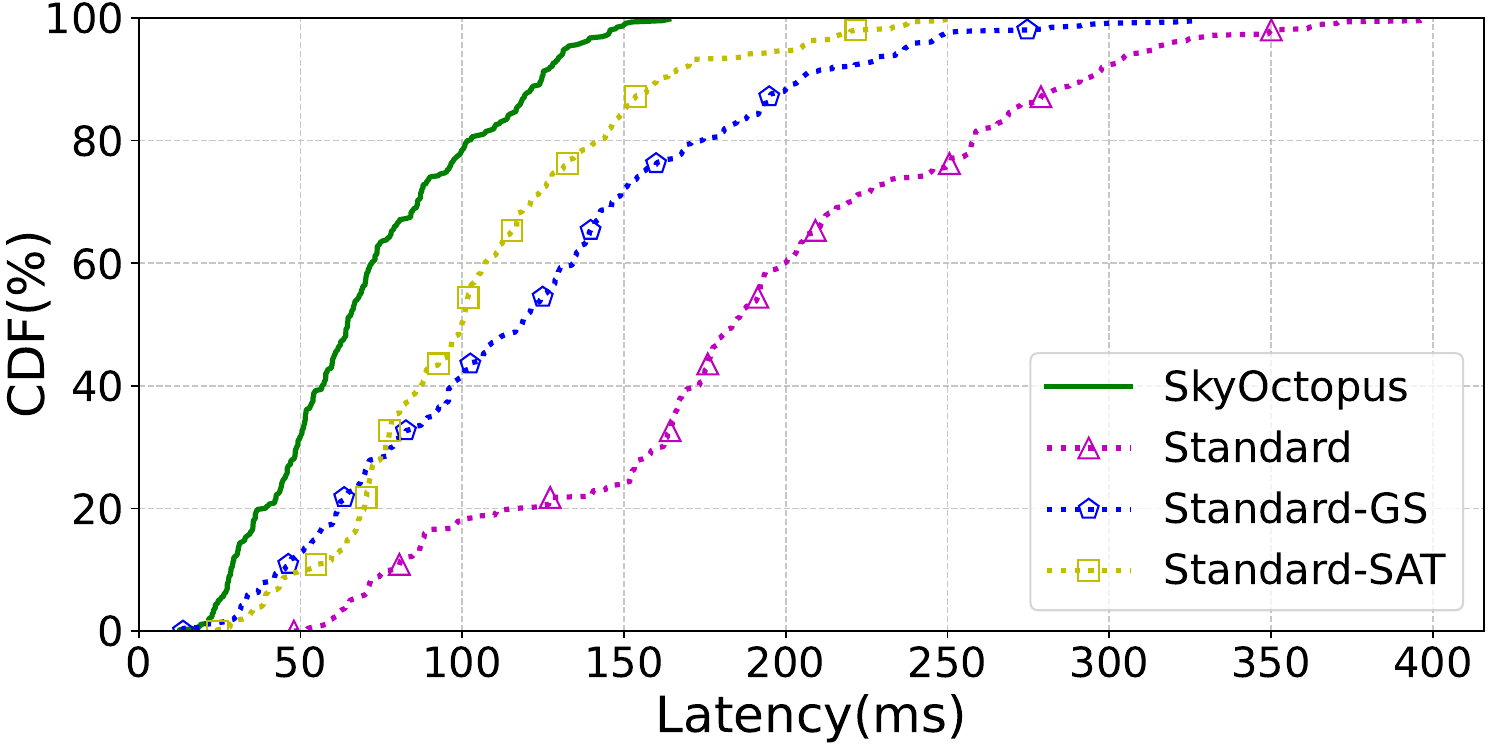}\label{first_starlink}}
    \subfloat[Kuiper]{\includegraphics[width=0.33\textwidth]{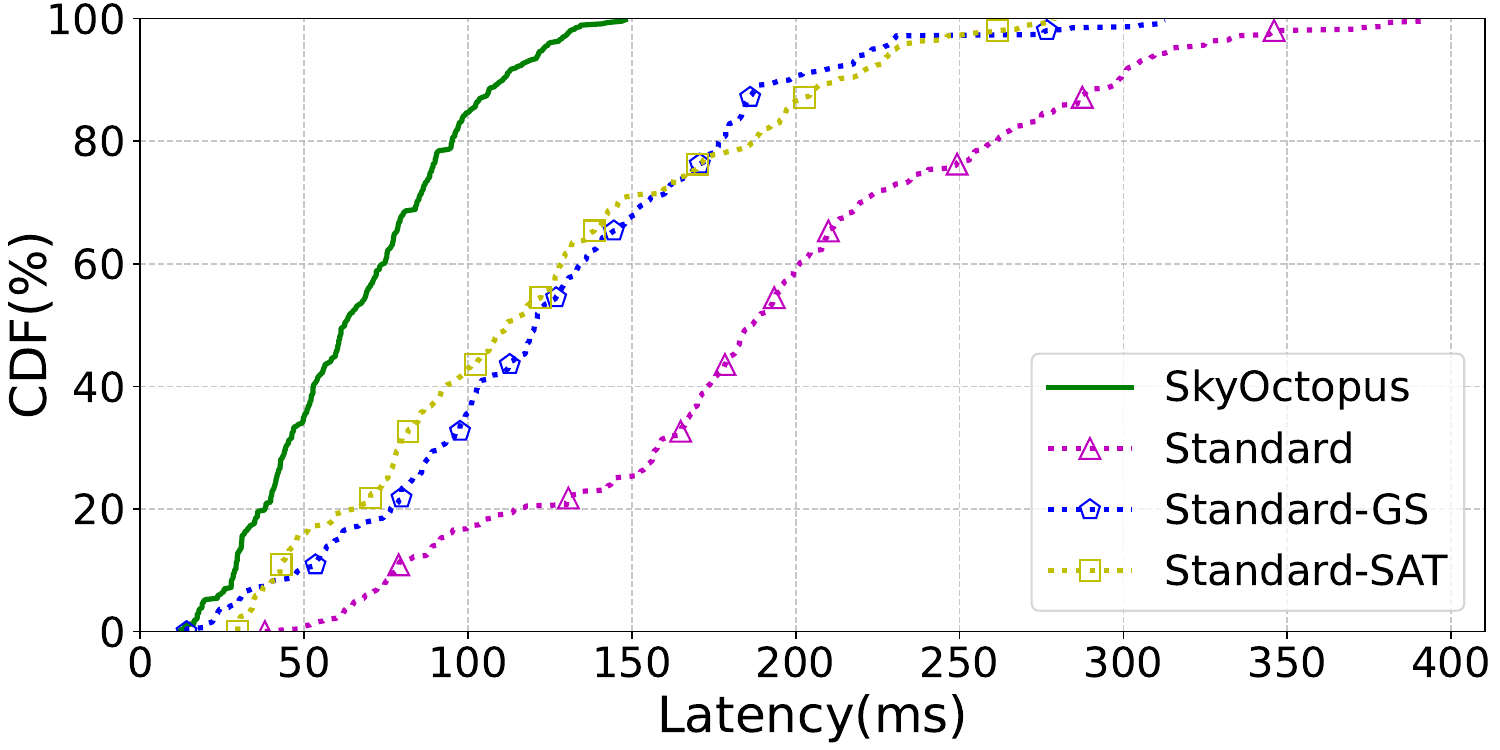}\label{first_kuiper}}
    \subfloat[OneWeb]{\includegraphics[width=0.33\textwidth]{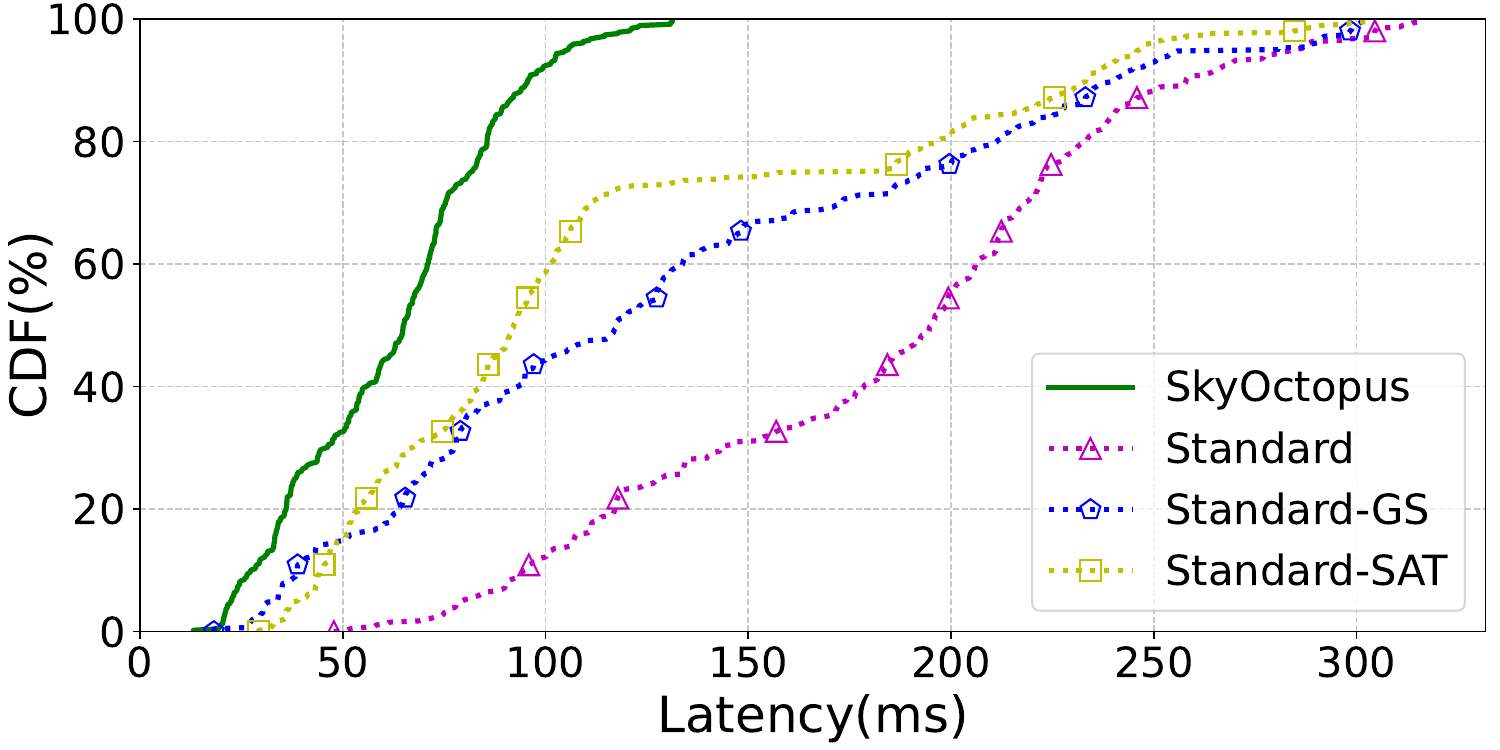}\label{first_oneweb}}
    \vspace{-0.2cm}
    \caption{Comparison of end-to-end latency in different constellations.} \label{latency_all}
    \vspace{-0.5cm}
\end{figure*}

\subsection{Experimental Results} \label{results of Experiment}


\noindent \textbf{End-to-end Latency:} We compare the end-to-end latency of different schemes using different constellations, which is shown in Fig.~\ref{latency_all}. It can be observed that SkyOctopus shows superior performance in terms of end-to-end latency compared to the three other schemes in different constellations. 

Specifically, Fig.~\ref{first_starlink} shows the latency in the Starlink constellation. The end-to-end latency based on SkyOctopus is 70.5ms on average, which is much shorter than the latency of 187.7ms based on the Standard scheme.  Meanwhile, it is much shorter than the latency based on the two other schemes (i.e., Standard-GS and Standard-SAT), which are 120.8ms and 104.5ms, respectively. This performance can be attributed to the fact that SkyOctopus can select the optimal anchor point from multiple paths, while the other three schemes always use the path corresponding to the fixed anchor point.

Furthermore, SkyOctopus's maximum end-to-end latency is 163.8ms, which saves 59\% and 50\% compared to 396.6ms of the Standard schemes and 327.9ms of the Standard-GS, respectively. This indicates that SkyOctopus can significantly reduce end-to-end latency for users in scenarios with severely circuitous routing.

We also evaluate the performance of SkyOctopus in other satellite constellations such as Kuiper and OneWeb, as demonstrated in Fig.~\ref{first_kuiper} and Fig.~\ref{first_oneweb}. SkyOctopus outperforms the other three schemes regardless of the type of satellite constellation. SkyOctopus primarily reduces latency by increasing the number of available anchor points, making it independent of the specific constellation. Besides, the latency exhibits a little difference in different constellations (6\% on average). This can be attributed to differences in constellation configurations, such as inter-orbit and intra-orbit distances and satellite altitudes. Although OneWeb has the highest satellite altitude, it benefits from a regular constellation topology, resulting in a lower average transmission latency in the satellite network. Therefore, it performs better than the other two constellations with the same ground station distribution.

Overall, SkyOctopus reduces end-to-end latency by an average of 53\% compared to the other three schemes in different satellite constellations.

\begin{figure}[]
	\centering
	\includegraphics[width=0.8\linewidth]{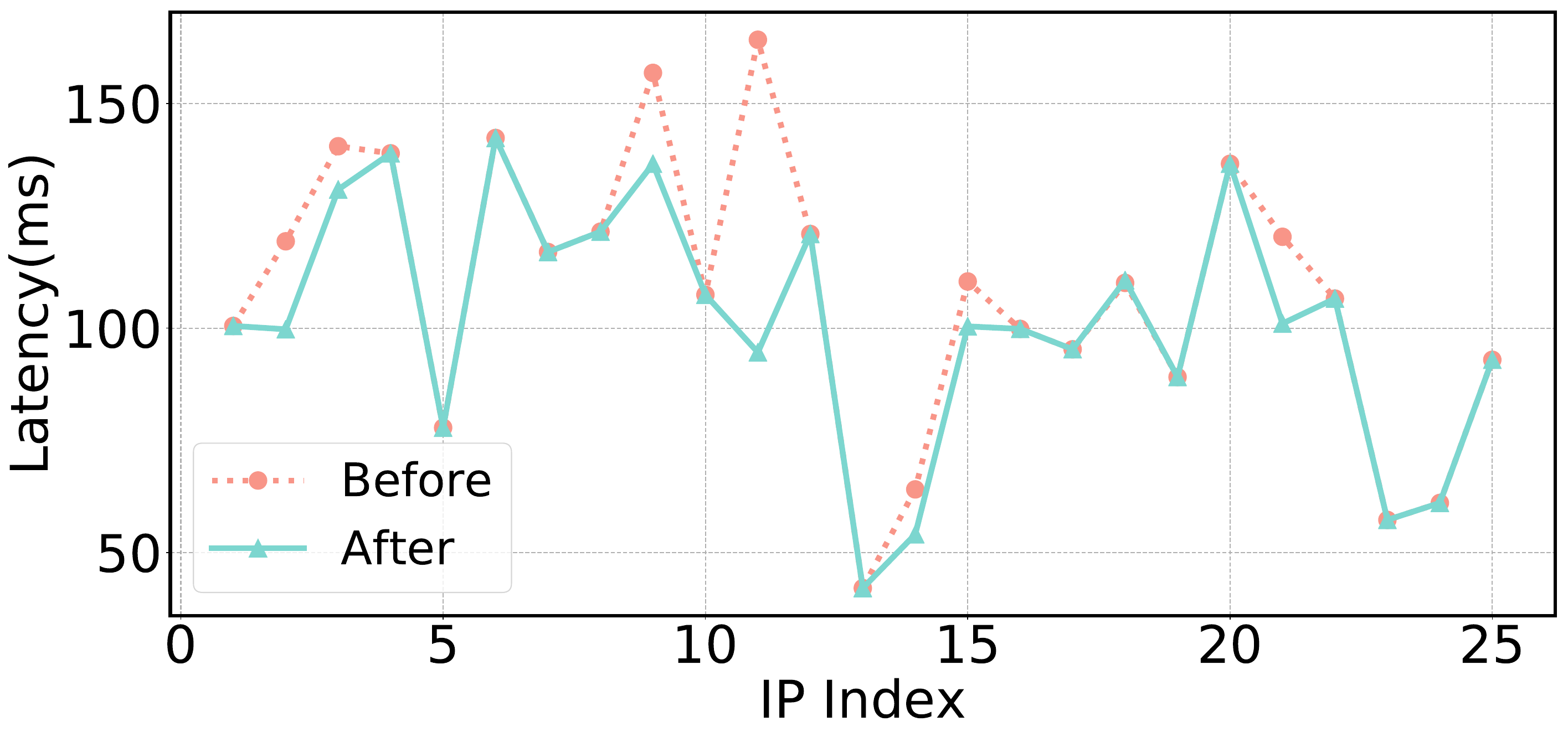}
        \vspace{-0.2cm}
	\caption{End-to-end latency to different target servers before and after path update. 
 }
	\label{anchor_selection}
 \vspace{-0.5cm}
\end{figure}

\noindent \textbf{Anchor Point Selection:} We recorded the latency experienced by one of the users when accessing these servers through SkyOctopus in the Starlink constellation to demonstrate the effectiveness of our anchor point selection strategy.

As shown in Fig.~\ref{anchor_selection}, we can see that the initially chosen anchor point based on location-based anchor selection has a 68\% concordance rate, meaning that the chosen anchor point after the first path update matches the initially chosen anchor point. In the remaining 32\% of non-concordance cases, the S-UPF promptly reselects anchor points for the user through the path update mechanism. In this situation, the user's latency decreases by a maximum of 42\% and an average of 21\%.

The reasons for these non-concordance cases primarily include three factors. The first reason is that, although the S-UPF forwards user traffic to the anchor point closest to the target server, the shortest distance between the anchor point and the target server does not always mean the lowest end-to-end latency. On the one hand, distance is only one of the factors affecting latency. On the other hand, in some cases, the intra-network latency may be significantly higher than inter-network latency. In such scenarios, the inter-network latency represents a small proportion of the end-to-end latency. Therefore, forwarding user traffic to the anchor point closest to the target server does not always result in the lowest end-to-end latency.

The second reason is the inaccurate correspondence between IP addresses and locations. This inaccuracy is influenced by the used database and its update frequency, as the accuracy of the database is challenged by various factors, such as dynamic IP address allocation and the use of proxy servers.

The third reason is the changes in network conditions along the path. Factors such as the movement of satellites, link congestion, and equipment failures can increase end-to-end latency. Some of these factors are difficult to predict, thus we introduce the path update mechanism to adjust PDRs timely. 

In summary, SkyOctopus can effectively select anchor points for user traffic and make timely adjustments to mitigate increases in end-to-end latency due to various factors.                                                                    

\begin{figure}[t!]
 \centering
    \begin{minipage}[t]{1\linewidth}
        \centering
        \vspace{-10pt}
        \subfloat[Starlink]{\includegraphics[width=0.33\linewidth]{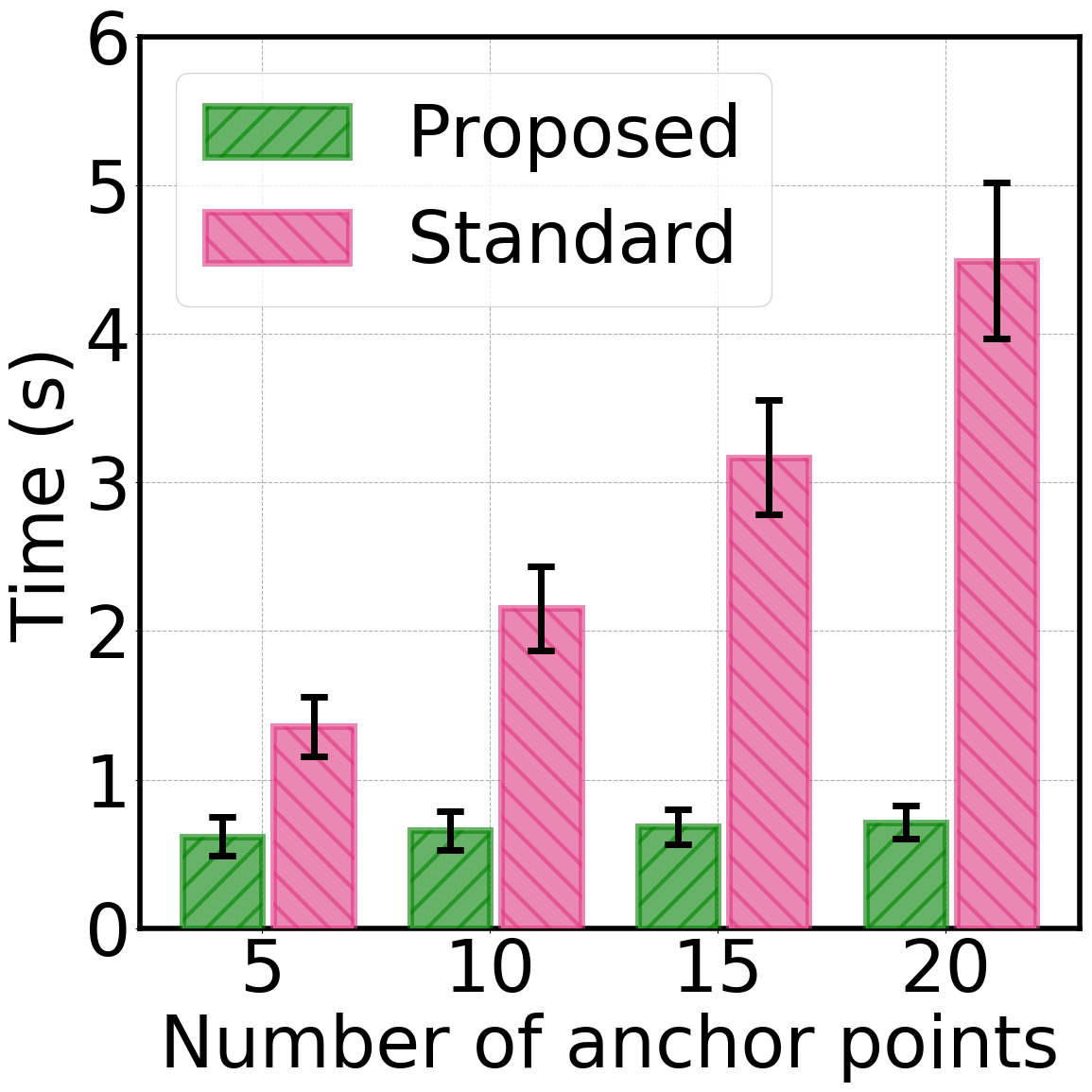}\label{session establishment starlink}}
        \subfloat[Kuiper]{\includegraphics[width=0.33\linewidth]{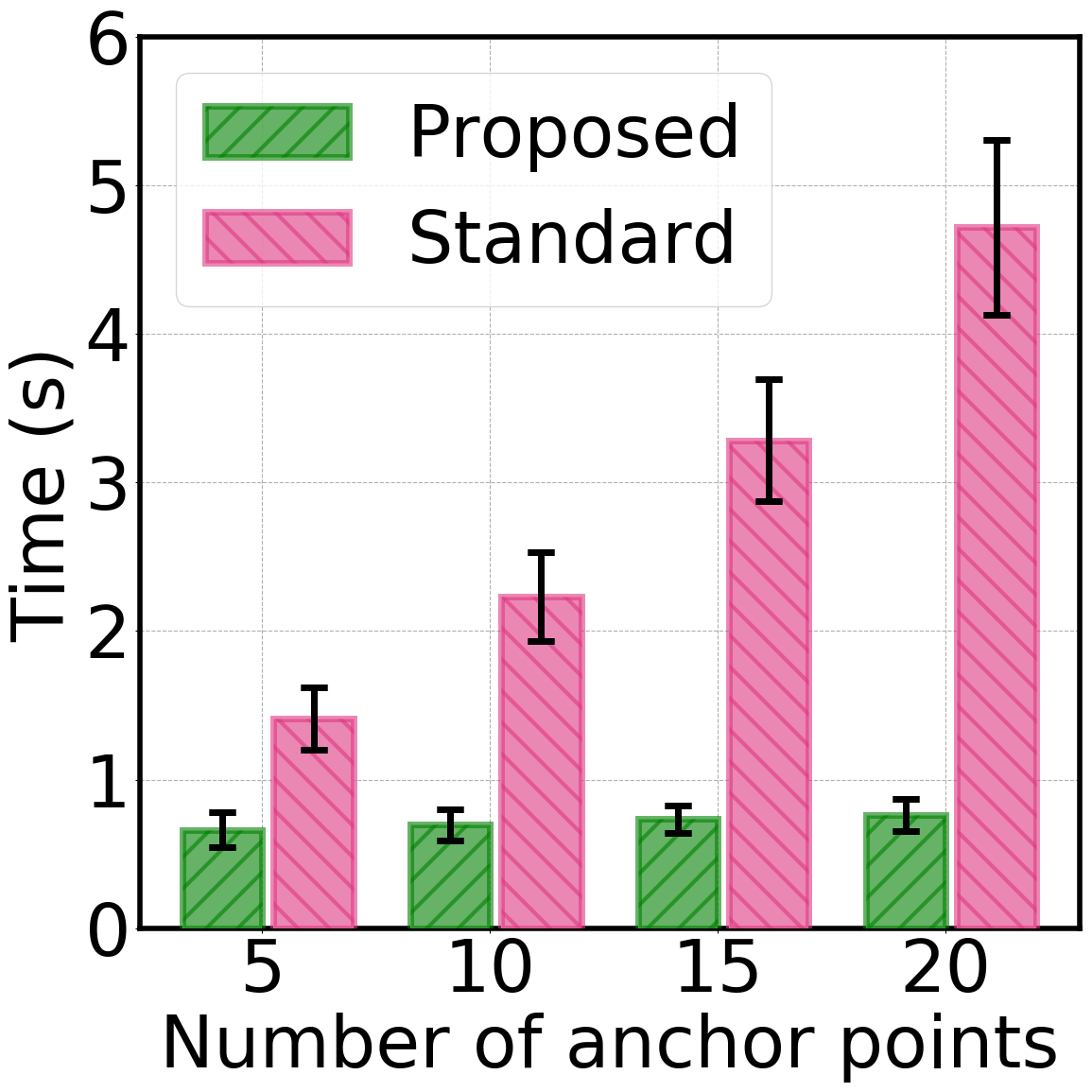}\label{session establishment kuiper}}
        \subfloat[OneWeb]{\includegraphics[width=0.33\linewidth]{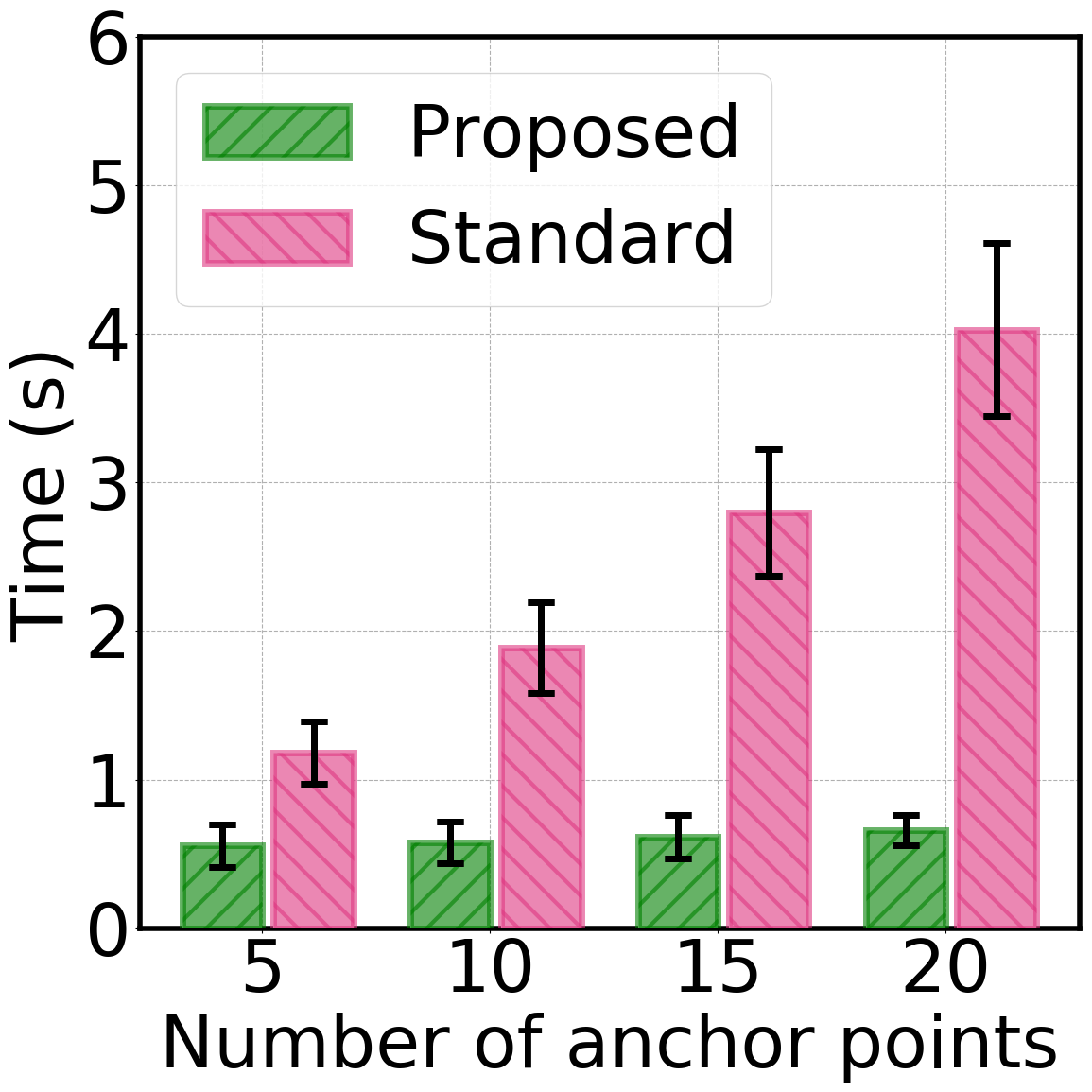}\label{session establishment oneweb}}
        \vspace{-0.2cm}
        \caption{Session establishment time in different PDU session establishment processes.}
        \label{session establishment}
        \flushleft
        \vspace{-0.2cm}
        \hspace{0.04\linewidth} \subfloat{\includegraphics[width=0.8\linewidth]{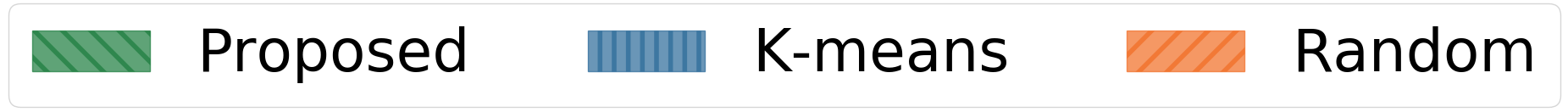}}
        \vspace{-0.35cm}
        \subfloat[Starlink]{\includegraphics[width=0.33\linewidth]{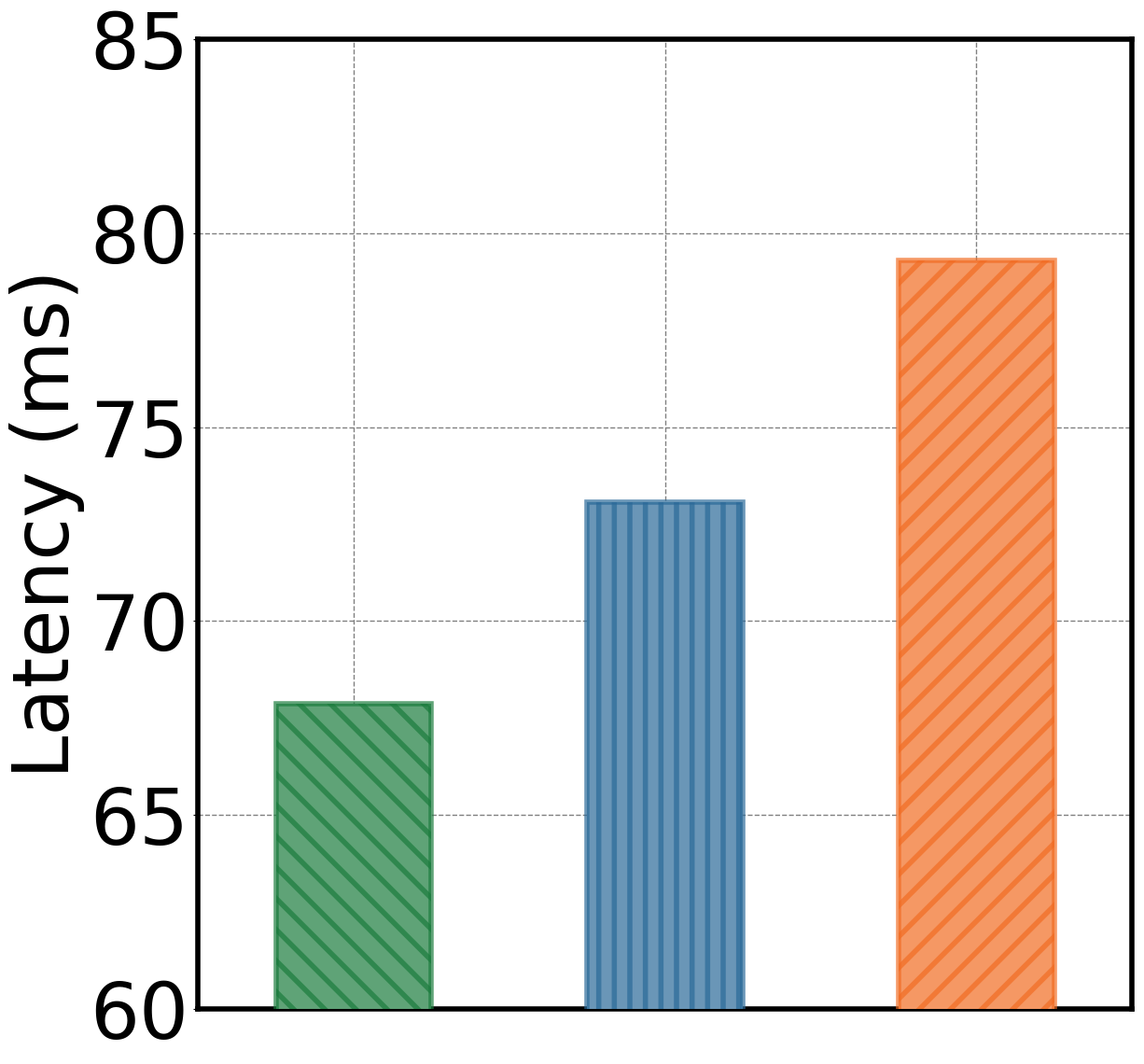}\label{anchor_distribution starlink}}
        \subfloat[Kuiper]{\includegraphics[width=0.33\linewidth]{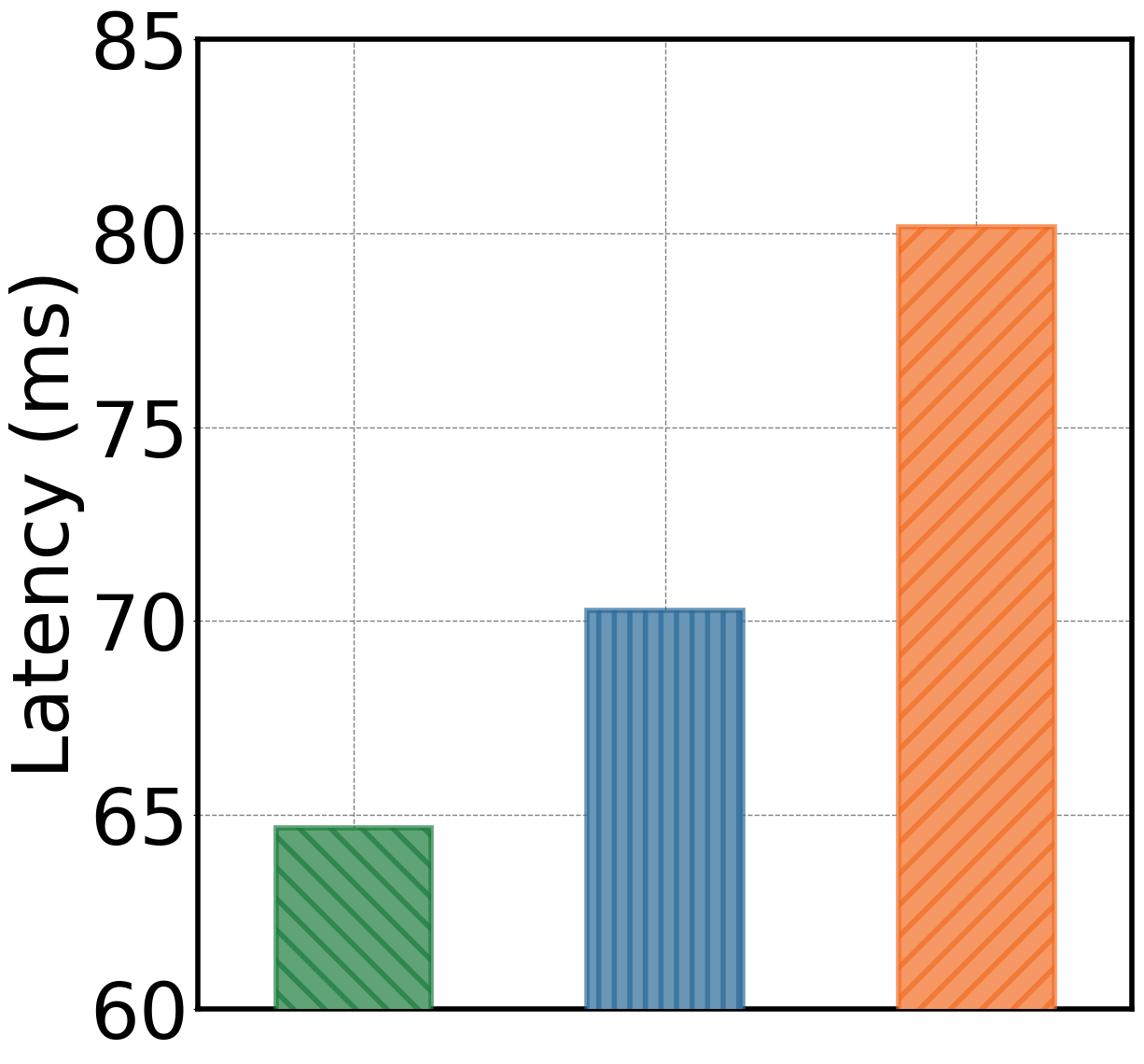}\label{anchor_distribution kuiper}}
        \subfloat[OneWeb]{\includegraphics[width=0.33\linewidth]{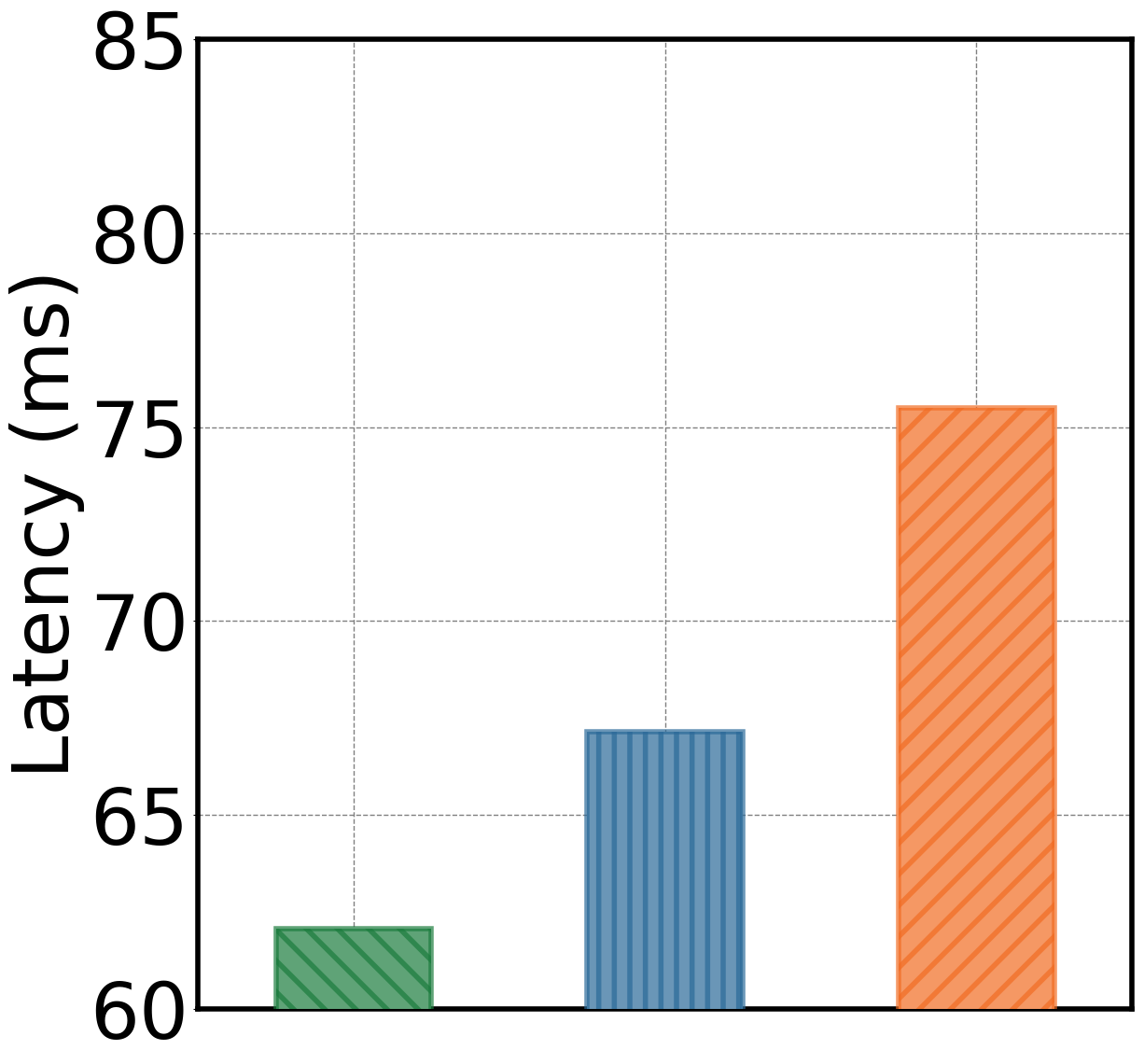}\label{anchor_distribution oneweb}}
        \vspace{-0.2cm}
        \caption{Performance of anchor point distribution algorithms.}
        \label{anchor_distribution}
    \end{minipage}
 \vspace{-0.5cm}
\end{figure}

\noindent \textbf{PDU Session Establishment Time:} Fig.~\ref{session establishment} shows the average time of users using different PDU session establishment schemes with different numbers of anchor points and constellations.
The proposed scheme saves 86\% of the average time with 20 anchor points compared to the standard 3GPP NTN scheme in all constellations. Additionally, the number of anchor points has little impact on the proposed scheme (with a maximum time variation of about 13\%), while the average time required by the standard scheme almost increases linearly as the number of anchor points grows.

Taking Starlink as an example (Fig.~\ref{session establishment starlink}), since the standard scheme always performs the insertion of each anchor point based on an already established PDU session, the interaction time of each anchor point with the core network is included in the total time, leading to an average time of 4.5s to establish a PDU session using the standard scheme with 20 anchor points. The proposed scheme achieves simultaneous interactions between the core network and multiple anchor points, meaning the time required to establish a session is mainly influenced by the longest time of all interactions, rather than all of them. Therefore, with the same number of anchor points, it only requires 713ms to establish the PDU session.

Meanwhile, the additional communication process between the S-UPF and the SMF in the proposed scheme runs in parallel with the communication process between the SMF and the base station. In most cases, the base station always establishes a connection with the S-UPF on the same satellite, and thus the time to the core network is nearly the same for both, which is another significant reason why the proposed scheme surpasses the standard scheme.

\noindent \textbf{Anchor Point Distribution:} Furthermore, we compare the impact of different algorithms on anchor point distribution in terms of average network latency. As shown in Fig.~\ref{anchor_distribution}, the proposed algorithm consistently achieves anchor point locations with lower average latency in various constellations. In the Starlink constellation, the proposed algorithm achieves an average latency of 67ms, while the K-means and random algorithms result in latencies of 73ms and 79ms, respectively. In the Kuiper and OneWeb constellations, there are slight differences in latencies. For the Kuiper constellation, the latencies for the proposed, K-means and random algorithms are 64ms, 70ms, and 80ms, respectively. For the OneWeb constellation, the corresponding latencies are 62ms, 67ms, and 75ms. On average, our algorithm reduces latency by 8.3\% compared to the K-means algorithm and by 17.2\% compared to the random algorithm.

We attribute this improvement to the fact that our proposed algorithm directly targets the optimization objective to find the optimal distribution. In contrast, the K-means algorithm attempts to find an anchor distribution that is as uniformly distributed as possible on the Earth's surface. However, a uniform distribution does not necessarily lead to lower latency due to the uneven distribution of target services and the non-linear relationship between geographical distance and latency.

\section{Related Work} \vspace{-0.5ex}
Since 2017, 3GPP has been continuously releasing technical specifications and reports regarding the integration of satellites and 5G~\cite{38811,38821,23737,28841,36763,38863}, sparking widespread discussion in the academic community about mobile satellite networks~\cite{discussion1,discussion2,discussion3}. Due to recent breakthroughs in technology and cost control for LEO satellites, as well as their inherent advantages over higher orbit satellites, many efforts have been made to integrate LEO satellites with mobile networks.

One mainstream approach is to modify the network structure to accommodate the movement of LEO satellites~\cite{spaceMobileNet, UPFandGNB, CoreDesign, anotherApp, satelessCoreDesign}. This includes discussing the deployment locations of network functions and introducing new network functions. However, most of these works do not focus on user plane issues or fail to provide sufficient improvements in reducing latency.
A recent work~\cite{spaceMobileNet} aims to shorten control plane latency by deploying part of the core network functions on satellites. However, it overlooks user plane issues, resulting in users still experiencing long end-to-end latency. 

Another approach attempts to overcome satellite mobility from a higher perspective by introducing new anchor management mechanisms without modifying the mobile networks themselves~\cite{skycastle, MM2, MM3}. These efforts often focus only on the latency variations caused by satellite mobility, rather than considering the entire end-to-end path. Work~\cite{skycastle} proposes a global mobility management mechanism, which provides low-latency global internet service to users through an anchor manager and distributed satellite anchor points. However, this mechanism focuses on latency changes within the satellite network and can only allocate anchor points that meet latency requirements rather than the optimal anchor point.

Our proposed architecture involves deploying the network function (i.e., S-UPF) on satellites. However, by redesigning the PDU session establishment process, we do not introduce additional overhead on the control plane. On the other hand, by expanding the available anchor points and comprehensively considering multiple end-to-end paths, we achieve a significant reduction in end-to-end latency, surpassing existing schemes. We consider both the user plane and the control plane and conduct comprehensive system-level experiments on a real data-driven platform, ensuring that the experimental results closely reflect real-world scenarios.

\section{Discussion and limitation} \vspace{-0.5ex}

\subsection{Satellite Computing}
With advancements in the space industry, deploying computational tasks on satellites is becoming feasible. By providing computational capabilities on satellites, mobile satellite networks will be able to offer MEC services to users with lower latency and reduced transmission rates. These services include a variety of applications, such as artificial intelligence (AI), remote sensing data processing, and the Internet of Things (IoT). SkyOctopus can be generalized to satellite computing, thereby enabling the offloading of user traffic to satellite computing nodes via the S-UPF.

\subsection{Content Delivery Network} 


In this paper, we consider that users always access the specific target server. However, the content delivery network (CDN) allows users to be served by the nearest available server. In such cases, SkyOctopus still provides users with the optimal anchor point selection, but the services requested by users may cluster around a specific anchor point, leading to an increase in its utilization. We believe that cooperation among anchor points can better adapt to CDN and leave it for future work.

\section{Conclusion} \vspace{-0.5ex}
Circuitous routing is a significant challenge faced by mobile satellite networks, which is largely caused by the design of a single-anchor session. In this paper, we propose an advanced mobile satellite network architecture. We introduce SkyOctopus, which avoids circuitous routing through multi-anchor connections and achieves low-latency access for users with a fine-grained anchor point selection strategy. Furthermore, we construct a prototype of SkyOctopus and conduct experiments to evaluate its performance. The results show that our solution can reduce end-to-end latency by up to 53\% and session establishment time by 86\%.

\bibliographystyle{IEEEtran}
\bibliography{newref.bib}

\end{document}